\documentclass[prl,floatfix,preprint]{revtex4}
\usepackage{graphicx}
\usepackage{epsfig}

\def  \bnabla  {\mbox{\boldmath$\nabla$}}

\begin{document}

\title{Strong electron spin-Hall effect by a coherent optical potential.}
\author {E. Ya. Sherman$^{1,2}$, J. G. Muga$^{1}$, V. K. Dugaev$^{3,4}$
and A. Ruschhaupt $^{5}$}

\affiliation{
$^1$Department of Physical Chemistry, Universidad del Pa\'is
Vasco UPV-EHU, 48080, Bilbao, Spain \\ 
$^2$  Basque Foundation for Science IKERBASQUE, 48011, Bilbao, Spain \\
$^3$Department of Physics, Rzesz\'ow University of Technology,
Powsta\'nc\'ow Warszawy 6, 35-959 Rzesz\'ow, Poland \\
$^4$Department of Physics and CFIF, Instituto Superior T\'ecnico, TU Lisbon,
Av. Rovisco Pais 1049-001 Lisbon, Portugal \\
$^5$ Institute for Theoretical Physics, Leibniz University  of Hannover,
Appelstr. 2, 30167 Hannover, Germany}

\date{\today}

\begin{abstract}
We demonstrate theoretically that a coherent manipulation of electron spins 
in low-dimensional semiconductor structures with spin-orbit 
coupling by infrared radiation is possible. 
The proposed approach is based on using a dipole 
force acting on a two-level system in a nonuniform
optical field, similar to that employed in the design 
of the cold atoms diode. For ballistic electrons the spin-dependent  
force, proportional to the intensity 
of external radiation, leads to a spin-Hall effect and resulting spin separation
even if the spin-orbit coupling itself does not allow for these effects.
Achievable spatial separation of electrons with opposite spins can 
be of the order of several tenth of a micron, an order of magnitude larger than
can be produced by the charged impurity scattering in the diffusive regime. 
\end{abstract}


\maketitle

\section{Introduction}

The development of abilities of manipulating spins
of itinerant and confined electrons is the central problem in modern
spintronics \cite{Dyakonov08}. A variety of techniques, not based on the magnetic field
effects, has been proposed for solving this problem in nonmagnetic
semiconductors. The first set of approaches employes the low-frequency
electric field applications. These approaches utilize
spin-orbit coupling, which couples spin to momentum, and, it turn, to the
low-frequency electric field which influences the electron motion.
Spin-orbit coupling here plays a role of a magnetic field acting on the electron
spin and spin-splitting the electron states energies \cite{Rashba03}. The other set of techniques,
described as the optical spintronics, is based on the indirect spin
manipulation by light with the photon energies of the order of one or
tenth of meV. The interest in the optical techniques is increasing because
they do not require electrical contacts attached to the sample and can allow
a very fast spin control. The simplest example is absorption of circularly
polarized light in a III-V semiconductor producing electron spin polarization \cite{Meier84}. 
One of the central concepts in spin manipulation is the spin current,
where electrons with opposite spins move in the opposite directions,
leading to spin accumulation at different regions of the sample. 
A conventional way to produce a spin current is based on the spin-Hall
effect induced by a constant electric field, where the sign of the Hall angle depends
on the electron spin direction \cite{Dyakonov08a}. A set of recent proposals for producing spin current,
accompanied or not by a charge current, 
is based on the coherent control of the one-photon and
two-photon processes and light absorption in quantum wells \cite{Bhat00,Stevens02,Najmaie03,Hubner03,Stevens03,Ganichev03,Bhat04,Najmaie05_1,
Sherman05,Tarasenko05}.

Almost all the present techniques for optical spin
manipulation are based on light absorption, that is on an incoherent process.
The only exception is the recently proposed stimulated Raman scattering
\cite{Najmaie05_2}, which can be used without photon absorption, being in this since, a
coherent process.

In atomic physics, spin-orbit and magnetic field effects can be introduced
by the coherent manipulations without incoherent interlevel transitions \cite{Sinova09}.
Here we propose purely coherent way of spin manipulation and spin current
production based on the force
acting on a two-level system close to the resonance in a nonuniform optical
field \cite{Metcalf}. These techniques are widely used in atomic physics to manipulate the
atomic motion but have not been considered for spins of electrons in solids
and spintronics applications. Here we propose a tool to inject a spin
current and spatially separate electrons with opposite spins by a technique
applied in optics to filter atoms in the ground and excited
states \cite{Muga1,Reizen09}. We will show that by a similar technique 
one can produce a spin-dependent dipole force acting on
electrons in two-dimensional semiconductor structures.

\section{Two-level system in resonant field: application to spins} 

To explain the physics of the spin separation, assume first that we have a general
two-level system with the ground $\left|g\right>$ (energy $E_{\left|g\right>}$) and excited $\left|e\right>$
(energy $E_{\left|e\right>}$) states. The system interacts  with electromagnetic radiation at the frequency $\omega$
close to the interlevel resonance at $\omega_{21}\equiv(E_{\left|e\right>}-E_{\left|g\right>})/\hbar$. 
In the rotating wave approximation, the Hamiltonian can be written as:
\begin{equation}
H=\frac{\hbar }{2}\left( 
\begin{array}{ll}
0 & \Omega _{R} \\ 
\Omega _{R} & -2\delta
\end{array}
\right) , 
\label{Hamilt}
\end{equation}
where $\Omega_{R}$ is the Rabi frequency, which we assume here for simplicity to be real,  
$\Omega_{R}$= $|\mathbf{dE}_{0}|/\hbar$, 
determined by the amplitude of oscillating external field $E_{0},$ $\mathbf{d}$ is
the transition dipole matrix element, and $2\delta=\omega-\omega_{21}$ is the frequency shift
with respect to the interlevel resonance.  We assume that the broadening of the
levels can be neglected compared to $\Omega_{R}$.  In the case of relatively weak external
field if $\Omega_{R}\ll|\delta|,$ the system is close to the ground state
$\left|g\right>$. The corresponding energy shift of the ground state obtained from Hamiltonian (\ref{Hamilt}): 
\begin{equation}
\Delta\varepsilon\simeq\frac{\hbar }{4}\frac{\Omega _{R}^{2}}{\delta }, 
\label{potential}
\end{equation}
depends on the sign of $\delta $. Assume that the electric field is
nonuniform, and, therefore, $\Omega _{R}^{2}$ depends on the position of the
two-level system. In this case, a dipole force \cite{Metcalf}  $\mathbf{F}=-\mathbf{\bnabla}(\Delta\varepsilon)$ acts
on the system. The direction of the force:
repulsion or attraction depends on the sign of $\delta,$ that is on the
shift of the radiation frequency with respect to the resonance.

Now we apply this approach to the spin-split electron states is quantum
wells. As an example we consider the (110) GaAs quantum well where the
spin-orbit coupling Hamiltonian has the form: 
\begin{equation}
H_{\rm so}^{\left[ n\right] }=\alpha ^{\left[ n\right] }k_{x}\sigma _{z},
\label{Hso}
\end{equation}
and we use the usual convention for the system axes with
respect to crystal axes as $x\parallel [1\overline{1}0],$ $y\parallel
[001],$ and $z\parallel [110]$. The coupling constant \cite{Dyakonov86} depends on the size quantization subband $n$ as: 
$\alpha ^{\left[ n\right] }=-\alpha _{D}n^{2}\left( \pi /w\right) ^{2}/2,$
where $\alpha _{D}$ is the bulk Dresselhaus constant, and $w$ is the quantum
well width. Therefore, the spin up $\left( \sigma _{z}=1\right) $-spin down $%
\left( \sigma _{z}=-1\right) $ splitting $\Delta E_{\mathrm{so}}^{\left[
n\right] }=2|\alpha^{\left[ n\right]}k_{x}|$ strongly depends on the
subband  number. The advantage of this well geometry is that the $%
z-$component of the spin is conserved even if momentum changes, and,
therefore, it relaxes very slowly. 

\begin{figure}[tbp]
\epsfxsize=5.5cm \epsfbox{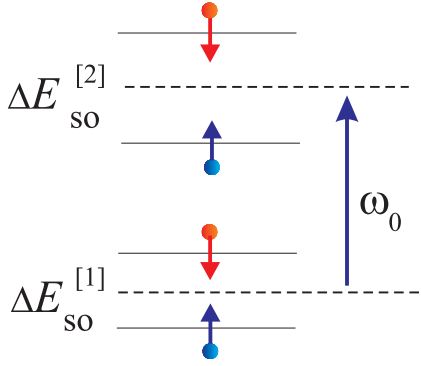}
\caption{Scheme of the spin-split energy levels; the radiation 
frequency $\omega_{0}$ is the distance between the size quantization subbands.}
\end{figure}

We consider now a quantum well irradiated by infrared light in the $p-$%
polarization mode at a nonzero incidence angle $\theta$ with the electric field
vector located in the incidence plane. As a result, the Poynting vector of the
radiation has components both in the plane of the well and perpendicular
to it. In this geometry nonzero $E_{z}$ 
in the well is produced at the assumed frequency $%
\omega_{0}$ exactly matching the intersubband resonance such that $\hbar\omega_0
=3\hbar ^{2}\left(\pi/w\right)^{2}/2m,$ where $m$ is the electron
effective mass. For simplicity we assume zero boundary conditions for the electron
wavefunction. In this
case, the transition between the spin up states is shifted with respect to
the resonance by 
$\left(\alpha ^{\left[2\right]}-\alpha^{\left[1\right]}\right)k_{x}=3\alpha^{\left[ 1\right] }k_{x},$ while the transition
between spin-down states is shifted by $-3\alpha ^{\left[ 1\right] }k_{x},$
as can be seen in Fig. 1.  Taking into account that typical values of $\Delta E_{\mathrm{so}}$ are of the
order of 1 meV, and the intersubband distance $\hbar\omega_{0}$ is of the order
of 100 meV, the required frequency stability  
of the radiation source should
be better than 0.1 percent. The broadening of the spin levels in this case can be neglected, too. 

According to Eq.(\ref{potential}), a spin-dependent
dipole force is acting on the electron in the case of coordinate-dependent electric
field, and, therefore, coordinate-dependent $\Omega_{R}$ defined as: 
\begin{equation}
\Omega _{R}=\frac{e}{\hbar }E_{z}\int_{-w/2}^{w/2}\psi _{2}(z)z\psi
_{1}(z)dz=\allowbreak \frac{16}{9\pi ^{2}}\frac{weE_{z}}{\hbar},
\end{equation}
where $\psi _{2}(z)=\sqrt{2/w}\sin (2\pi z/w)$ and $\psi _{1}(z)=\sqrt{2/w}%
\cos (\pi z/w)$ are the corresponding eigenfucntions. This approach is
implemented in the proposal for the atomic diode \cite{Muga1,Reizen09}, where the field is
produced by a laser beam close to the atomic resonance.

To estimate the upper boundary of the potential
and to understand whether it is possible to observe the spin separation
experimentally, we 
begin with the estimate of electric field $E_{c}$ at which
the Rabi frequency $\Omega_{R}$ is of the same order of magnitude as the SO
splitting  $\Delta E_{\mathrm{so}}^{\left[ 1\right] }/\hbar $: $\hbar \Omega
_{R}\sim E_{c}ew\sim \Delta E_{\mathrm{so}}^{\left[ 1\right] }.$ 
With $\Delta E_{\mathrm{so}}^{\left[ 1\right] }\sim 1
$ meV, $w\sim10$ nm, one obtains $E_{c}\sim 10^{3}$ $\mathrm{%
V/m.}$ The corresponding absolute value of the Poynting vector $\left\langle S_{c}\right\rangle
=\epsilon _{0}cE_{c}^{2}/2\sim 0.1$ $\mathrm{MW/m}^{2}$ is not high and can
be achieved experimentally. 

\section{Spin separation for electrons moving through a microchannel} 

To demonstrate the spin separation effect, we consider a spin-unpolarized flow of electrons
through a channel of a micron or submicron width \cite{Matsuyama02} 
extended along the $x$-axis and
irradiated by an electromagnetic wave in the $p-$mode. This field has a component
perpendicular to the quantum well and, therefore, leads to a nonzero Rabi
frequency. We assume for simplicity that all electrons entering the channel 
have the same momentum $k$ and velocity $v=\hbar k/m$ along the $x$-axis. 
 The electrons with almost parallel momenta can be injected from a two-dimensional
electron gas through a tunneling barrier (as shown in Fig.2) or through
a narrow collimator. The tunneling probability strongly depends 
on the component of momentum along the channel and, therefore, leads
to the collimation of the outgoing electrons. We concentrate below at the electrons that can pass through the
channel.  
The resulting optical spin-dependent potential $U_{\sigma }(x,y)$ and 
$y-$component of the force $F_{\sigma}(x,y)$, as follows from Eqs.(\ref{potential}) and
(\ref{Hso}) are:
\begin{equation}
U_{\sigma }(x,y)=\frac{\hbar^2}{6}\frac{\Omega_{R}^{2}\left( x,y\right) }{%
\Delta E_{\mathrm{so}}^{[1]}}\sigma_{z},\text{\qquad}F_{\sigma }(x,y)=-%
\frac{\partial U_{\sigma }(x,y)}{\partial y},
\end{equation}
where for brevity we use $\sigma=\pm 1$ instead of $\sigma_z$. We consider only the $y$-component of the force since it leads to
a qualitative effect: spin current across the channel and spin accumulation
at its boundaries. The $x$-component of the force, $-{\partial U_{\sigma }(x,y)}/{\partial x}$
is nonzero, too. However, it only slightly accelerates electrons along the channel,
and, despite a spin-dependence, does not have a qualitative effect 
on the spin density. The coordinate dependence of the Rabi frequency is given by the field
distribution at the quantum well surface 
\begin{equation}
\Omega _{R}^{2}\left(x,y\right) =\Omega _{R}^{2}\left( 0\right) \exp \left[
-\left( y^{2}/\Lambda _{y}^{2}+x^{2}/\Lambda _{x}^{2}\right) \right] ,
\end{equation}
where $\Lambda_{x}$ and $\Lambda _{y}$ are the spot dimensions of the order of
light wavelength of tens of microns. The parameters  $\Lambda_{x}$ and $\Lambda_{y}$
depend on the shape of the infrared beam and its incidence angle. Our following
analysis does not depend on the details of the field distribution provided it is 
inhomogeneous both in the $x$ and $y$-directions. In addition,
we assume that the channel boundaries, even if modify the irradiating  field
distribution, do not distort its shape strongly. 

The microchannel is shifted at the
distance $y_{0}$ of the order of $\Lambda _{y}$ from the spot center to
assure a nonvanishing dipole force. We also assume that the channel width $d$ is
much smaller than $\Lambda_x$ and $\Lambda_y$  such that the available $y$ are limited by: $%
y_{0}-d/2<y<y_{0}+d/2.$ The resulting optical potential can be estimated as $%
U_{\sigma }(x)\sim \pm 0.1$ meV. The corresponding force $F_{\sigma }(y)\sim
\left(\hbar\Omega_{R}^{2}/\Delta E_{\mathrm{so}}^{[1]}\Lambda_{y}\right)
\sigma_{z}$ leads to the spatial transverse separation in the velocities
and coordinates: spin-up (spin-down) electrons are accumulated at the upper
(lower) boundary of the channel, as shown below.

\begin{figure}[tbp]
\epsfxsize=5.5cm \epsfbox{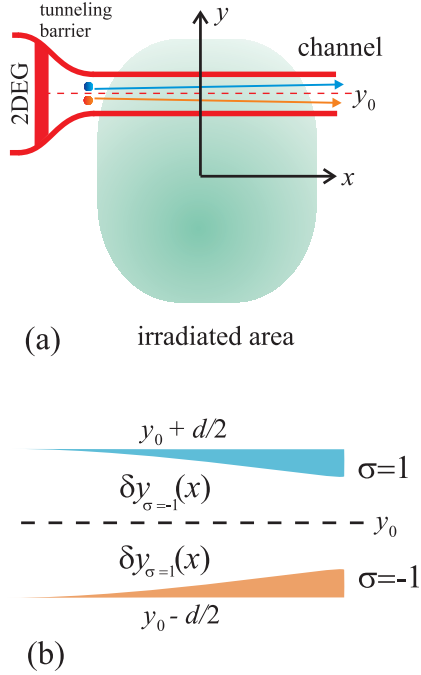}
\caption{(a) Separation of spin-up and spin-down electrons moving through the irradiated channel.
Injection of collimated electrons can occur from a two-dimensional electron gas (2DEG) reservoir
through a tunneling barrier. We assume that in the 2DEG 
the second subband of the size quantization is empty. 
Regarding the channel, we assume that
the initial concentration of electrons there is very low, such that in the steady state
it is due to the carriers passing through the channel from the 2DEG.
(b) Schematic plot of the spin accumulation (darker areas) in the channel,
according to Eqs.(\ref{n:minus}) and (\ref{n:plus}).
The positions of the boundaries of the spin accumulation areas 
with respect to the channel boundaries are $\delta y_{\sigma}(x)$
($\delta y_{\sigma=1}(x)=-\delta y_{\sigma=-1}(x)$). We note that 
here $\delta y_{\sigma=-1}(x)<0$.}
\end{figure}

In the irradiated area, the electrons with opposite spins are
accelerated in the opposite directions, producing spin-Hall effect
in a coherent optical potential. A nonzero spin current $%
K_{y}^{z}\left( x,y\right) $ for $\sigma _{z}$ spin component propagating in
the $y-$direction: 
\begin{equation}
K_{y}^{z}\left( x,y\right) =\sum_{\sigma}n_{\sigma }\left( x,y\right)
v_{y}^{\sigma }\left( x,y\right),
\end{equation}
appears,  where $n_{\sigma }\left(x,y\right)$ is the spin-projected density, 
and $v_{y}^{\sigma}\left(x,y\right)$ is the corresponding
velocity component with 
$v_{y}^{\sigma=-1}\left(x,y\right)=-v_{y}^{\sigma=1}\left(x,y\right)$. 
We assume that in the geometry shown in Fig.2, all electrons
reaching the channel boundaries, are absorbed there.  
The spin up electrons are accumulated near 
the upper side of the channel, electrons with spin down - at the lower side of it. 
As a result, a spin density pattern is formed.  
We consider electrons moving through the irradiated channel, where the force $F_{\sigma }(y)$ is
approximately uniform across it and only the linear terms can be
taken into account. In this case the acquired velocity and displacement  
for electrons, which have not yet reached the boundaries  are
given by: 
\begin{eqnarray}
&&v_{y}^{\sigma}(x)= \frac{1}{m}\int_{-\infty}^{x}F_{\sigma }(l,y_{0})\frac{dl}{v}
=\frac{1}{m}\int_{-\infty }^{x}F_{\sigma }(0,y_{0})\exp\left(-l^{2}/\Lambda_{x}^{2}\right)\frac{dl}{v}, \label{vysigma}\\
&&\delta y_{\sigma}(x)=\int_{-\infty }^{x}v_{y}^{\sigma}(l)\frac{dl}{v}.
\label{ysigma}
\end{eqnarray}
The first equality in Eq.(\ref{vysigma}) is exact and does not depend on the details of the
incident field  distribution. Solutions of equations (\ref{vysigma}) and (\ref{ysigma}) 
have the asymptotic behavior: $v_{y}^{\sigma}(x\gg \Lambda _{x})=F_{\sigma }(0,y_{0})%
\sqrt{\pi}\Lambda _{x}/mv,$ and $\delta y_{\sigma }(x\gg \Lambda _{x})=v_{y}^{\sigma }(x\gg
\Lambda _{x})x/v.$ The resulting spin accumulation near the boundaries has the form:
\begin{eqnarray}
n_{\sigma=-1}\left( x,y\right) =\frac{n}{2},\quad n_{\sigma=1}\left( x,y\right)=0 \qquad {\rm if } \quad y-\left( y_{0}-\frac{d%
}{2}\right)\le \delta y_{\sigma=1}(x), \label{n:minus}\\
n_{\sigma=1}\left( x,y\right) =\frac{n}{2},\quad n_{\sigma=-1}\left( x,y\right)=0  \qquad {\rm if } \quad y-\left( y_{0}+\frac{d%
}{2}\right)\ge \delta y_{\sigma=-1}(x), \label{n:plus}
\end{eqnarray}
as shown in Fig. 2(b), and $n$ is the total concentration of electrons in the channel
in the steady state.  If  neither of the spatial conditions in Eqs.(\ref{n:minus}), (\ref{n:plus})
is satisfied, then $n_{\sigma=-1}=n_{\sigma=1}$, and the total spin density is zero. 
The transverse
spin-dependent velocity change $v_{y}^{\sigma}(\Lambda_{x})$ can be estimated
as $F_{\sigma }t_{\Lambda}/m\sim \hbar \Omega _{R}^{2}/mv\Delta E_{\mathrm{%
so}}^{[1]},$ where $t_{\Lambda }=\Lambda _{x}/v$ is the irradiated channel
part passing time, and does not depend strongly on the spot size. Corresponding
spin-dependent transverse separation of electrons passed through 
the channel  $\delta y_{\sigma}(\Lambda _{x})$ can be
estimated as: $v_{y}^{\sigma}(\Lambda _{x})t_{\Lambda }\sim \left(
U/\varepsilon \right) \Lambda _{x}.$ With $L\sim 10$ $\mu $m$\sim 10^{-3}$
cm, $U_{\sigma }/\varepsilon\sim 1/30$ we obtain $\delta y_{\sigma}(\Lambda_{x})\sim 0.3$
$\mu $m as an upper estimate of the displacement. The resulting spin deviation angle of the order of 
$\delta y_{\sigma}(\Lambda _{x})/\Lambda_{x}\sim 3\times 10^{-2}$ 
is an order of magnitude larger than the spin-Hall angle due
to the spin-dependent scattering by charged impurities \cite{Huang2004}
in the diffusive regime. In the ballistic regime, the Hamiltonian in Eq.(\ref{Hso}) 
does no lead to a transverse spin separation even if a static external 
electric field is applied along the channel. Therefore, in the absence
of external radiation, spin-Hall effect in the system we consider, will vanish.    

In the above consideration we assumed ballistic motion of the 
electrons through the channel. This assumption is justified by the
fact that at mobility $10^{6}$ cm$^2$/Vs achieved  in
high quality samples and electron velocity
$10^{8}$ cm/s, the free path for GaAs is close to 35 micron, much larger than the spot
size. 

\section{Conclusion}

We have shown that electron spins in low-dimensional
semiconductor structures can be manipulated coherently by
coordinate-dependent optical fields at infrared frequencies, producing
optical form of spin-Hall effect for electrons. 
The resulting spin-dependent forces are similar to the atomic state-dependent 
dipole forces in nonuniform optical fields used to manipulate motion of cold atoms.
The separation of ballistically moving electrons with opposite spins, 
vanishing in the absence of external radiation, can be of the order of few tenth 
of a micron, that is much larger than can be expected for the spin-Hall effect due 
to the scattering by the charged impurities in the diffusive regime. 
Our model can be naturally extended to more general spin-orbit coupling
Hamiltonians, including the Rashba term \cite{Rashba84}.

\textit{Acknowledgement}

E.S. and J.G.M. are grateful to the support of University of Basque Country EPV-EHU 
(Grant GIU07/40), Basque Country Government (IT-472-10), and  Ministry of 
Science and Innovation of Spain (FIS2009-12773-C02-01). 
V.D. is partly supported by the FCT Grant PTDC/FIS/70843/2006 in Portugal and by 
Polish Ministry of Science and Higher Education research project in years 2007 -- 2010.


\newpage

\begin{thebibliography}{99}


\bibitem{Dyakonov08}  {\it Spin Physics in Semiconductors}
                      Springer Series in Solid-State Sciences, Vol. 157, 
                      Dyakonov  M I  (Ed.), 2008,  442 p.

\bibitem{Rashba03}  Rashba  E I and  Efros Al L
                    2003 {\it Phys. Rev. Lett.}  \textbf{91} 126405;
                    2003 {\it Appl. Phys. Lett.} \textbf{83} 5295

\bibitem{Meier84}  \textit{Optical Orientation}, 
                   edited by  Meier F and Zakharchenya  B P,  
                   Modern Problems in Condensed Matter Sciences, Vol. \textbf{8}
                   (North-Holland, Amsterdam, 1984).

\bibitem{Dyakonov08a} Dyakonov  M I and Khaetskii  A V
                     in Ref.\cite{Dyakonov08}, p. 210.

\bibitem{Bhat00}    Bhat R D R and Sipe J E 
                    2000 {\it Phys. Rev. Lett.} \textbf{85} 5432

\bibitem{Stevens02} Stevens  M J,  Smirl A L, Bhat R D R, Sipe J E and van Driel H M 
                    2002 {\it J. Appl. Phys.}  \textbf{91} 4382

\bibitem{Najmaie03} Najmaie  A, Bhat R D R and Sipe J E  
                    2003 {\it Phys. Rev.} \textbf{B68} 165348.

\bibitem{Hubner03}  H\"{u}bner J, R\"{u}hle  W W, Klude  M,  Hommel D, Bhat R D R, Sipe J E and van Driel H M
                    2003 {\it Phys. Rev. Lett.}   \textbf{90} 216601 .

\bibitem{Stevens03}  Stevens  M J,  Smirl A L, Bhat R D R, Najmaie A, Sipe J E and van Driel H M
                     2003 {\it Phys. Rev. Lett.}  \textbf{90} 136603 

\bibitem{Ganichev03}   Ganichev S D and Prettl W
                      2003 {\it  J. Phys. Cond. Matt.} \textbf{15} R935, and reference therein.

\bibitem{Bhat04}  Bhat R D R, Nastos F, Najmaie A and Sipe J E
                  {\it Phys. Rev. Lett.} {\bf 94} 096603 (2005) 

\bibitem{Najmaie05_1}  Najmaie A, Smirl A L and Sipe J E 
                     2005 {\it Phys. Rev.} \textbf{ B71} 075306 .

\bibitem{Sherman05}  Sherman E Ya, Najmaie A and Sipe J E
                     2005 {\it Appl. Phys. Lett.}  \textbf{86} 122103 

\bibitem{Tarasenko05}  Tarasenko S A  and Ivchenko  E L 
                       2005 {\it Pis'ma v ZhETPh}  \textbf{81} 292; 
                       Ivchenko E L and Tarasenko S A
                       2008 {\it Semicond. Science and Techn.}   {\bf 23} 114007

\bibitem{Najmaie05_2} Najmaie A, Sherman E Ya and Sipe J E
                    2005 {\it Phys. Rev. Lett.}  \textbf{95} 056601;
                     Najmaie A, Sherman E Ya and Sipe J E,
                     2005 {\it Phys. Rev.}  \textbf{B72} 041304

\bibitem{Sinova09} Liu  X J,  Borunda M F,  Liu X and Sinova J
                    2009 {\it Phys. Rev. Lett.}  {\bf 102} 046402 

\bibitem{Metcalf}  Metcalf H J and van der Straten P 
                   {\it Laser Cooling and Trapping} Springer (1999).

\bibitem{Muga1}  Ruschhaupt A and Muga J G 
                  2006 {\it Phys. Rev.} {\bf A73} 013608;
                  Ruschhaupt A and Muga J G 
                  2004 {\it Phys. Rev.} {\bf A70} 061604;
                  Ruschhaupt A, Muga J G,  and Raizen M G
                  2006  {\it Journ. of Phys. B: Atomic Molecular and Optical Physics}  {\bf 39} L133 

\bibitem{Reizen09}  Raizen M G 
                   2009 {\it Science}  {\bf 324} 1403 

\bibitem{Dyakonov86}   Dyakonov M I and  Kachorovskii V Yu 
                       1986 {\it Sov. Phys. Semicond.} \textbf{20} 110; 
                       for holes:  Rashba E I and Sherman  E Ya
                       1988 {\it Phys. Lett.} \textbf{A129} 175 

\bibitem{Matsuyama02}  Matsuyama T,  Hu C M, Grundler D, Meier G and Merkt U
                       2002 {\it Phys. Rev.} {\bf B65} 155322 

\bibitem{Huang2004}  Huang H C, Voskoboynikov O and  Lee C P
                     2004 {\it J. Appl. Phys.} {\bf 95} 1918 

\bibitem{Rashba84}  Bychkov Yu A and  Rashba E I
                    1984 {\it JETP Lett.} \textbf{39} 79;
                    Rashba E I
                    1964 {\it Sov. Phys. - Solid State} \textbf{2} 1874


\end{thebibliography}
\end{document}